\documentstyle[prl,tighten,aps]{revtex}

\begin{document}

\title{Inference of Planck action constant by a classical 
fluctuative postulate holding for stable microscopic
and macroscopic dynamical systems} 

\author{Salvatore De Martino$^{a,b,}$\thanks{E--Mail: 
demartino@vaxsa.csied.unisa.it},
Silvio De Siena$^{a,b,}$\thanks{E--Mail: 
desiena@vaxsa.csied.unisa.it} 
and Fabrizio Illuminati$^{a,b,c,}$\thanks{E--Mail: 
illuminati@vaxsa.csied.unisa.it} \\
{\em $^a$Dipartimento di Fisica,}\\ 
{\em $^b$INFN, Sez. di Napoli and INFM Unit\`a di Salerno, \\
Universit\`a di Salerno, I-84081 Baronissi (SA) Italy,} \\
{\em $^c$Fakult\"{a}t f\"{u}r Physik, Universit\"{a}t Potsdam, \\
     Am Neuen Palais 19, D--14469, Potsdam, Germany.}}
\date{4 August 1998}
\maketitle

\begin{abstract} 
The possibility is discussed of inferring or
simulating some aspects of quantum dynamics 
by adding classical statistical fluctuations
to classical mechanics.   
We introduce a general principle of mechanical stability
and derive a necessary condition for classical chaotic
fluctuations to affect confined dynamical systems,
on all scales ranging from microscopic to macroscopic 
domains.   
As a consequence we obtain, both for microscopic
and macroscopic aggregates, dimensional relations 
defining the minimum unit of action of individual 
constituents, yielding in all cases Planck 
action constant. 
\end{abstract}

\vspace{0.2cm}


\section{Introduction}

Since the origin of quantum mechanics, the question of its 
relationship to classical mechanics has been recognized
to be crucial for a deeper physical understanding of
the theory.

The WKBJ analysis of the
semiclassical limit and the coherent states of
the harmonic oscillator are classic examples
that show all the subtleties, both mathematical
and conceptual, of the classic to quantum interplay.

The issue has recently raised widespread interest
on account of remarkable experimental advances in the
study of mesoscopic and macroscopic quantum phenomena 
such as high $T_{c}$ superconductivity, atom cooling
and Bose--Einstein condensation, and the coherent
quantum--like states of charged beams in particle
accelerators.
At the same time, for some theoretical
problems such as quantum computing and the quantization
of gravity, semiclassical methods have so far been the
main or sometimes the only tool that led to some partial
results.

Traditionally, the main theoretical tools employed to recover
classical aspects of quantum systems and study the
transition between the classical and the quantum domain
fall in two classes: on the one hand 
asymptotic series expansions in powers of $\hbar$, such
as the semiclassical WKBJ approximation; 
on the other hand, the singling out of particular
quantum states exhibiting some degree of space and time 
coherent behavior.

In all such cases the arrow points in the direction
going from the deep quantum domain heading on to recover
classical aspects of the dynamics in some suitable limit.

In the present paper we show, in a still semi--quantitative 
framework, that it is possible to take the opposite route, 
namely to infer fundamental quantum aspects 
by starting from classical mechanics 
and then adding suitable, purely classical
fluctuations to the classical deterministic motion:
these fluctuations then simulate or describe
some features of the fundamental quantum noise.

Remarkably, one needs to assume only a minimum set of very
general properties about the fluctuations, the analysis of
their detailed structure being, at this stage, unnecessary.
In particular, they are assumed to be universal (i.e. 
independent of the systems and of the interactions considered),
and to act coherently on all length scales (``coherent
tremor hypothesis''). The scale of action of these fluctuations
then naturally turns out to be, in all cases
considered, the Planck constant. Most remarkable
is that this occurence happens to be true also
for classical macroscopic systems whose dynamics
is commonly described in the framework of classical
mechanics.
A preliminary announcement containing
partial results appeared in \cite{noilettera}.

The starting point for our analysis is a scheme 
recently proposed by Francesco Calogero \cite{calogero}.
In his reasoning Calogero suggests that the origin of 
quantization be attributed to the universal interaction of every
particle with the background gravitational force due to all other 
constituents of the Universe, which should result in some effects
of chaoticity and stochasticity on sufficiently small
space--time scales.

Assuming such a noisy component of the motion
to be a coherent tremor in the sense we have 
explained above, and assuming a basic
granularity of the Universe, made up of nucleons of mass $m$ 
interacting via the gravitational force, Calogero derives an expression
for Planck's constant, $h=G^{1/2} m^{3/2} R^{1/2}$, with $G$ the
Newtonian gravitational constant and $R$ the observed radius of the
Universe. This formula, which connects the fundamental constant of
quantum theory with the fundamental gravitational constants, was
already known since some time \cite{weinberg}, but its meaning
and implications were so far unexplained.

In the present paper we reformulate the scheme of Calogero
dropping the hypothesis of the gravitational chaos, and
replacing it with the assumption of a general principle
of stability for classical dynamical systems. 
The requirement of mechanical stability holds
independently of the interaction being considered
(as long as the latter is attractive), and allows us
to derive the coherent tremor hypothesi of Calogero,
rather than postulating it.

The method can then be applied, besides gravity,
to all other interactions, for instance 
electromagnetic and strong forces.
In this way we are able to show that for any bound
and stable aggregate of particles interacting 
through any (overall attractive)
phenomenological law of force, 
the coherent tremor hypothesis yields  
precise dimensional relations between the
minimum unit of action per particle, 
the fundamental constants associated to the 
given interaction, and the dimension (radius) of 
stability of the given aggregate.

In order of magnitude, such dimensional relations
always yield Planck action constant $h$, and this 
is the main result of the present work.

The extension to finite temperature, and other
dimensional estimates that come as applications
of the method are also discussed in the  
paper, which is organized as follows.

After a brief review in Sect. 2
of the original model introduced by Calogero,
we reformulate in Sect. 3 the problem in a general 
form combining the classical laws 
of force with a simple principle of dynamic stability. 
The coherent tremor hypothesis of Calogero is then 
readily derived in a very natural way.

We then show in Sect. 4 that the coherent tremor 
hypothesis can be reformulated in a ``Keplerian'' version 
which is independent of the number of particles
(constituents) of the dynamical system under consideration. 
In the case of astrophysical systems associated to gravitational 
interactions, this form recovers already
known features of some long--range classical
cosmological scaling laws \cite{weinberg}.

In Sect. 5 we apply the method to all possible 
interactions (gravitational, electromagnetic, strong)
giving rise to stable bound systems of particles.
We find that, in each of such cases, the order of magnitude of  
the characteristic unit of action per particle (constituent)
equals that of Planck constant. 
Remarkably, we consider also the case of {\it macroscopic},
therefore substantially classical, aggregates of charged
particles like neutral plasmas and, particularly,
charged particle beams in accelerators. 
The method yields, even for these classical systems, a unit
of action per constituent of the order of $h$.
Estimates on
the phenomenological ranges of weak, screened electromagnetic
interactions for mesoscopic systems are then derived; for
electrostatic dipolar forces, we obtain that the unit of action
per particle is directly linked to the fine structure constant,
and we comment on this intriguing connection.

In Sect. 6, a reasonable hypothesis on the form of the
characteristic time scale of the classical fluctuations,
already introduced by Calogero, combined with the natural
definition of a thermodynamic equilibrium energy allows us
to extend the procedure to finite temperature, and to check
the consistency of our scheme in several interesting cases.
In particular, we obtain a general expression for the
equivalent thermal unit of action, that scales in a 
nontrivial way as teh square root of the number of
constituents. Applyinf this definition to
the instance of charged particle beams we link 
the beam unit of emittance to  
the number of particles in the beam. Among other
quantities that we can estimate by our method, we have
reported the characteristic velocity of a condensed particle
in a Bose--Einstein condensate, and the order of magnitude
of the total number of nucleons in the Universe, in excellent
agreement with the estimates obtained by cosmological
arguments. 

Finally, in Sect. 7 we comment on the nature and meaning of
our results, and draw our (tentative) conclusions, expecially
by discussing about the possible future perspectives and 
directions of research along the line introduced in the 
present paper.

Before entering the detailed discussion, we warn the reader that
throughout all computations we have
neglected all numerical factors 
that do not affect the order of 
magnitude of the estimated quantities.

\section{The scheme of Calogero: the coherent tremor hypothesis}

The fundamental assumption in the scheme of Calogero \cite{calogero}
is that, due to the universal and long--ranged nature of the
gravitational interaction, to the granular structure of the
Universe (made of an exceedingly large number of microscopic atomic
constituents, for instance nucleons), 
and to the unavoidable presence of a chaotic component
of the motion entailed by classical mechanics for a large
many--body system, every and each particle in the Universe experiences
locally the same stochastic gravitational acceleration. Calogero
identifies in this phenomenon the universal stochastic
background responsible for quantization. 

Therefore, in his view, quantum mechanics should be derivable
by classical mechanics plus a suitable {\it ad hoc} assumption
on the nature and structure of the gravitational background
stochastic fluctuations: the crucial point
in the procedure is then the introduction of the
characteristic time scale $\tau$ of the local,
stochastic component of the motion for each single constituent.

Calogero assumes, rather naturally, 
that the characteristic frequency of the chaotic
motion must be larger, the larger the number $N$ of elementary
constituents in the Universe. 
Moreover, being this a collective stochastic
effect, he infers that the fluctuations should be
normally distributed and therefore scale 
proportionally to $\sqrt{N}$. 

Hence, he puts forward the estimate
\begin{equation}
\tau \cong N^{-1/2} {\cal{T}},
\end{equation}
where ${\cal{T}}$ is the characteristic scale of time 
for the mean global deterministic motion of each 
constituent in the Universe. Eq. (1) is the mathematical
expression of the universal tremor hypothesis.

He defines also the characteristic global units of 
energy $E$ and of action $A$ 
of the Universe, and the unit of energy per 
particle (constituent) $\epsilon$:
\begin{equation}
A \cong E {\cal{T}}\, ,
\end{equation}
\begin{equation}
\epsilon \cong E/N \, .
\end{equation}

\noindent Finally, he introduces the characteristic 
unit of action per particle $\alpha$,
defined by
\begin{equation}
\alpha \cong \epsilon \tau \, .
\end{equation}

The definitions (2), (3), (4) and the assumption (1) connect the unit of
action $\alpha$ to the total action $A$ through the nontrivial relation
\begin{equation}
\alpha \cong N^{-3/2} A \, .
\end{equation}

\noindent If $M$ is the total mass of the 
Universe, $R$ its observed radius and $G$
the universal gravitational constant, 
a simple dimensional analysis yields
\begin{equation}
A \cong G^{1/2} M^{3/2} R^{1/2} \, .
\end{equation}

\noindent 
Estimating the total number of particles by $N \cong M/m$, with $m$
the mass of a nucleon (or of a hydrogen atom), Calogero finally
obtains
\begin{equation}
\alpha \cong G^{1/2} m^{3/2} R^{1/2} \, .
\end{equation}

\noindent
Inserting the numerical values $m \cong 10^{-27} Kg$, $G \cong 10^{-11}
Kg^{-1} m^{3} s^{-2}$, and the most updated cosmological estimate for
the observed radius of the Universe $R \cong 10^{30} m$ \cite{peebles},
eq. (7) yields, in order of magnitude,
$\alpha \cong h$, i.e. Planck action constant.

This estimate leads Calogero to infer the congruence and the
cogency of his hypothesis, for classical gravity plus
the tremor hypothesis (1) allows him to {\it derive} Planck constant.
He provides also a semi--quantitative argument (based on a simple
nonrelativistic Newtonian model) to show 
that the above picture is consistent only 
with a spatially coherent character of the tremor,
in the sense that the (tiny) corrections to the 
relative displacements between particles due to the stochastic 
background must be proportional to the relative displacements
themselves, through a term which might depend on time. He refers 
to this large--scale coherence by the fascinating definition
``breathing of the Universe'' \cite{calogero}.

Concluding the section we wish to recall that the picture
of Calogero, together with his interpretation of it, is based on
the explicit consideration of the exceedingly large number of
microscopic constituents forming the Universe, 
which may justify invoking classical chaos as the founding 
stone of a possible gravitational stochastic background 
giving origin to quantization.

In the next sections we will see that hypothesis (1)
can be viewed as the consequence of a principle
of classical mechanical stability,
and that it leads to a characteristic unit of action per 
particle of the order of $h$, Planck constant, 
for any stable system, irrespective of the number of 
its constituents. 

\section{General formulation: the principle of stability, and the
derivation of the coherent tremor hypothesis}

In this section we generalize the scheme of Calogero to any 
bound, stable classical dynamical system of global dimension $R$
and global time scale ${\cal{T}}$, formed  
by $N$ elementary constituents.

We will show that the characteristic action per particle for any
aggregate can be connected to the related phenomenological force, 
and that from this procedure the tremor statement (1) follows in a
natural way.

Let us consider a system described by a classical law of force $F(r)$.
By introducing the mass $m$ of a generic elementary constituent, its
mean global characteristic velocity $v=R/{\cal{T}}$ and its
characteristic time scale $\tau$, 
we define the characteristic unit of action per particle as
\begin{equation}
\alpha \cong m v^{2} \tau \, .
\end{equation}

Now, an elementary criterion of dynamical stability for
classical bound systems requires that on average
the characteristic potential energy of a particle must be of the
same order of magnitude of its characteristic kinetic energy.
As a matter of fact, this principle reduces to the virial
theorem \cite{landau}, complemented with the hypothesis 
that the mean kinetic energy per particle and the
mean potential energy per particle do not depend on the
particular constituent being chosen.
Then, if ${\cal{L}}$ denotes the characteristic work performed 
on average
by the system on a generic constituent, one has to impose that
\begin{equation}
{\cal{L}} \cong m v^{2} \, .
\end{equation}

\noindent On the other hand, 
\begin{equation}
{\cal{L}} \cong N F(R_{m}) R_{m} \cong N F(R) R \, ,
\end{equation}

\noindent
where $R_{m}$ denotes some mean length scale, which is obviously of
the same order of magnitude of the global dimension $R$.

Eqs. (9), (10) allow to express the characteristic velocity as
\begin{equation}
v \cong \sqrt{N F(R) R m^{-1}} \, \, .
\end{equation}

\noindent Reminding the definition
\begin{equation}
v \cong R {\cal{T}}^{-1} \, ,
\end{equation}

\noindent we can put together the definition (8) 
with the expressions (11), (12),
to finally obtain for the characteristic unit of
action per particle the following expression:
\begin{equation}
\alpha \cong m \sqrt{N F(R) R m^{-1}} \frac{\tau}{{\cal{T}}}
= m^{1/2} R^{3/2} \sqrt{F(R)} \frac{\tau}{{\cal{T}}} \sqrt{N} 
\, \, .
\end{equation}

Looking at the expression (13), we can see, now, that the fluctuation
hypothesis of Calogero Eq. (1) can be immediately derived by imposing 
what we can denote as a
``statement on the universal stability of matter", namely by 
requiring that the characteristic unit of action per particle be 
independent from the number $N$ of elementary constituents.
In this way we obtain, besides the position (1) for the 
characteristic time 
$\tau$, the following form of the characteristic 
unit of action per particle:
\begin{equation}
\alpha \cong m^{1/2} R^{3/2} \sqrt{F(R)} \, ,
\end{equation}

\noindent
which connects $\alpha$ with the characteristic equilibrium
dimension of the system, its characteristic constants 
(such as the masses of the elementary constituents) and the
classical phenomenological law of force associated to it.

Application of relation (14) to specific constituents,
forces, and systems will then provide
the order of magnitude of
the characteristic unit of action per particle 
$\alpha$ for the various
possible physical stable aggregates.

We have thus far extended the scheme of Calogero into a 
more general theoretical framework.
Before turning to the applications of Eq. (14) to actual physical
systems in order to compute the unit of action per particle in several
cases of fundamental interest, we show in the next section
that the coherent tremor hypothesis (1) can be reformulated
in a form indipendent by the number of constituents $N$, 
allowing us to introduce a characteristic scale of length 
$l$ connected to the characteristic time $\tau$.

\section{General formulation: the Keplerian fractal form of the tremor}

We consider at first the particular case of the Universe.
In order to reformulate position (1) in a form which
is independent from the number $N$ of constituents, we define the 
mean volume per particle $V/N$, with $V$ the mean global volume of
the Universe, and a characteristic length
\begin{equation}
l \cong (V/N)^{1/3} \, .
\end{equation} 

\noindent The following simple chain of relations then
trivially follows from Eqs. (1) and (15):
\begin{equation}
\tau^{2} \cong \frac{V}{N} \frac{{\cal{T}}^{2}}{V} 
\cong l^{3} \frac{{\cal{T}}^{2}}{V} \cong l^{3} 
\frac{{\cal{T}}^{2}}{R^{3}} \, ,
\end{equation}

\noindent where we have exploited 
$V \cong R^{3}$, with $R$ the observed radius.
On the other hand, ${\cal{T}}$ is of the same order of 
magnitude of the period of a
particle subject to the third Kepler law and performing an
orbit of radius $R$. But exactly Kepler law assures that 
${\cal{T}}^{2}/R^{3}=const$, and therefore Eq. (16) yields
the relation
\begin{equation}
l \sim \tau^{2/3} \, ,
\end{equation}

\noindent where the symbol $\sim$ indicates proportionality.
As the exponent $2/3$ reminds of the large scale Kepler law, 
we refer to relation (17) as to the ``Keplerian" 
form of the tremor. The fractal exponent involved is a
signature of the fluctuative nature of the tremor hypothesis.

Considering now stable aggregates different from the Universe, 
we see that, due to the simple fact that these systems must have 
well defined volume $V (\cong R^{3})$ and characteristic unit
of time ${\cal{T}}$, it is trivially true that 
${\cal{T}}^{2}/R^{3}=const$, independently from the validity of
Kepler law. Then, the statement (1), the definition (15) and
the chain of relations (16) will lead again to relation (17).
Also in this case we will speak of ``Keplerian" form of
the tremor.

Some comments are now in order.
The definition (15) allows to interpret the characteristic length
$l$ as a ``mean free path" per particle, but also as the ``mean
position uncertainty" per particle; this last interpretation
could be more suitable as long as quantum aspects are involved.

As already pointed out, the term ``Keplerian form" for relation
(17) is not referred, in general, to some periodic 
orbit and a true Kepler 
law, but it is only reminiscent of the Kepler $2/3$ exponent.
The interesting features of the form (17) are its independence
from the number $N$ of constituing particles,
and its nonanalytic character which is typical of classical
stochastic fluctuations.

\section{Computation of the characteristic unit of action per
particle for generic stable dynamical systems}

We will now use expression (14) to compute the order of
magnitude of the characteristic unit of action per particle
associated to several stable aggregates of constituents 
interacting through different (overall attractive)
laws of force.

\vspace{0.5cm}

{5.1 {\it Gravitational interactions}}

In this case, we insert in eq. (14) the gravitational law of
force $F(R)=G m^{2}/R^{2}$ between pairs of nucleons
(hydrogen atoms) of mass $m \cong 10^{-27} Kg$, 
calculated at some average distance of the same order 
of magnitude of the global size of the system, i.e. the
observed radius of the Universe $R \cong 10^{30}$m.
The result is just the expression (7) for the quantum 
of action (Planck constant), originally derived by Calogero.

\vspace{0.5cm}

{5.2 {\it Electromagnetic interactions}}

This case requires a somewhat more accurate analysis with respect
to the gravitational one. To begin with, one can consider the most 
elementary stable aggregate of electromagnetic nature, provided by
the electrostatic interaction between two elementary charges of
opposite sign, for instance an electron and a proton in 
a hydrogen atom: $F=e^{2}/R^{2}$ (we use electrostatic units
throghout).
The basic relations (13) and (14) then yield
\begin{equation}
\alpha \cong m^{1/2} e R^{1/2} \, .
\end{equation}

\noindent Here $e \cong 10^{-14} N^{1/2}$m is the elementary charge,
while the mean interparticle distance, of the order of magnitude 
of the size of the system, is of course taken to be the
Bohr radius: $R \cong 10^{-10}$m; finally, for $m$ we
choose the reduced mass between electron and proton masses,
whose value is practically of the order of magnitude of the
electron mass $m \cong 10^{-30} Kg$.

Inserting these numerical values in Eq. (18) we obtain again 
$\alpha \cong h$, the Planck action constant.

The situation becomes even more intriguing 
when one considers more complex systems formed by many
charged particles subject to screening effects of the charge
distributions, and to continous processes
of photoemission and absorption. Electromagnetic interactions
exhibit a wide range of validity, which extends from the
microscopic scale of the hydrogen atom ($\cong 10^{-10}$m)
to the intermediate scales typical of large molecules and
molecular clusters ($\cong 10^{-6}$m), 
up to the macroscopic scales.

In the present context, one can take into account screening 
effects and the permanent emission and reabsorbtion of
photons by including in our treatment a new fundamental 
constant ruling these processes. For electromagnetic
interactions this cannot be else than
the velocity of light $c$. The computation of the 
characteristic action per particle becomes then dependent
on the phenomenological laws of force that can be built 
by taking combinations of powers not only of $e$, $m$, and
$R$, but also of $c$. We can take the exponents of the
powers of the former three quantities to be parametrically
dependent on the exponent, say $\gamma$, of the powers
of $c$:
\begin{equation}
F(e,m,R,c^{\gamma}) = (e^{2})^{a(\gamma)}m^{b(\gamma)}
R^{d(\gamma)}c^{\gamma} \, ,
\end{equation}

\noindent and then proceed to determine them for each
chosen value of $\gamma$. The value $\gamma = 0$ of
course yields Coulomb law (no screening effects, neglect
photoemissions and absorbtions). Negative values of
$\gamma$ define screened Coulomb forces, that decay
faster than $R^{-2}$. We now give some examples:

a) $\gamma = - 1$: in this case 
\begin{equation}
F = \frac{e^{3}}{m^{1/2}cR^{5/2}} \, \, ,
\end{equation}

\noindent and this leads to a unit of action per
particle that has the following expression:
\begin{equation}
\alpha = m^{1/4}e^{3/2}c^{-1/2}R^{1/4} \, \, .
\end{equation}

\noindent Due to the exponent $1/4$ the variation
of $R$ over a wide range of values does not
sensibly affect the order of magnitude of $\alpha
\cong h$; however, the equality is optimized for
$R \cong 10^{-6}$m, which corresponds to the size
of large molecules and molecular clusters.

b) $\gamma = -2$: in this case
\begin{equation}
F = \frac{e^{4}}{mc^{2}R^{3}} \, \, ,
\end{equation}

\noindent and this yields a unit of action
\begin{equation}
\alpha = \frac{e^{2}}{c} \, \, .
\end{equation}

\noindent We see that in this case the dependence
of $\alpha$ on the dimensions of the system disappears.
Moreover, the unit of action turns out to be the ratio
$\hbar/\alpha_{em}$, i.e. of Planck constant to the
electromagnetic fine structure constant. 

We have not got
at present a clear understanding of this very intriguing
result and of its possible deep significance. However, we
notice that the effective dipole--dipole force given by
eq. (22) is known to be responsible for the binding of
dipolar crystals. The latter can exist in chains of more or
less arbitrary length, ranging from atomic dimensions
(monocellular crystals), up to macroscopic ones, and
are moreover very stable. This observation in some
sense justifies the absence of the length scale $R$
of the aggregate in eq. (23). The latter relation can
then be interpreted as a crude approximation to a 
dependence given by a very small power: $\alpha \sim
R^{\delta}$, with $\delta$ very close to zero.
However, the appearance of the fine structure constant
and its relation to the fluctuative hypothesis and to
dipolar effective interactions is more intriguing
and deserves to be better understood. 

In conclusion, we have shown that by taking into account
screening effects and photoemission in some effective
way by inserting powers of $c$ in the determination
of the phenomenological law of force, we can move 
smoothly from the microscopic domain towards larger
scales and more complex structures, by requiring the
unit of action per particle to be fixed at the value
(in order of magnitude) of Planck constant. Viceversa,
fixing the dimensions of the mesoscopic and macroscopic
aggregates in the average expressions of the phenomenological
force laws, and inserting the latter in our expression for
the characteristic unit of action per particle, yields
(in order of magnitude) Planck constant.

However, in order to support our point of view, it is
crucial to test our scheme in the cases of truly
{\it macroscopic} charged aggregates, which are 
universally recognized, like in the case of gravitational
interactions, to be included in the classical
domain, and whose dynamics is commonly studied
in the framework of classical mechanics.
We move then to consider two relevant instances of
such systems: neutral plasmas and 
charged beams in particle accelerators.

The case of neutral plasmas \cite{plasma} is somewhat problematic.
First of all, these systems are only approximately
in a full equilibrium state (for instance one cannot
define a global equilibrium temperature, but
one can only attribute a different temperature for
each kind of charged particle species that is present). 
Since our method is explicitely designed to deal
with stable, bound systems, we cannot ``a priori"
expect such a spectacular appearance of the
Planck constant (in order of magnitude) for the
unit of action per particle in each specific
instance of neutral plasmas.

In fact, we cannot even define unambigously a
``global" characteristic dimension, 
in particular, a global coherence length.
However, we can consider data connected to the 
plasma oscillations, which in globally neutral
plasmas are described by an elastic effective force
\begin{equation}
F = - k_{p} x \, .
\end{equation}

\noindent acting on electrons \cite{plasma}.
Here, $x$ describes the distance inside a double stratus
of opposite charges, while the elastic constant $k_{p}$
is given by \cite{plasma}
\begin{equation}
k_{p} = 4 \pi n_{e} e^{2}\, ,
\end{equation}

\noindent with $n_{e}$ the number of electrons per unit of
volume and $e$ the electron charge $e \cong
10^{-14} N^{1/2}$m. Finally, the characteristic
linear dimension is given by the Debye radius $R_{D}$,
which represents the maximal elongation of the
oscillations.

We are then describing the neutral plasma as 
composed by subsystems (the oscillating strata)
which can be considered, to some extent depending 
on the particular plasma being considered, 
nearly mutually independent and in 
dynamical equilibrium.
Obviously, this description will be valid
only in some crude approximation.

We apply now relation (14) to these subsystems,
choosing for the mass that of an electron,
and for the characteristic dimension 
associated to stability the Debye radius. 
We obtain then
\begin{equation}
\alpha \cong m^{1/2} {k_{p}}^{1/2} {R_{D}}^{2} \, .
\end{equation}

\noindent At this point we shall consider that
one is concerned with a wide
variety of plasma systems: celestial plasmas 
(ionosphere, solar photosphere, interstellar matter),
gaseous discharge plasmas, thermonuclear fusion
plasmas, and so on, each associated to particular
values of the parameters \cite{plasma}.

As we have already remarked above,
we cannot expect to obtain $\alpha
\cong h$ in all such cases.
However, for high pressure and
extremely high pressure gaseous discharge plasmas
(for which $n_{e} = 10^{21}$m$^{-3}$,
$R_{D} = 10^{-7}$m and $n_{e} = 10^{23}$m$^{-3}$,
$R_{D} = 10^{-9}$m, respectively),
for plasmas in stationary regimes induced
by lasers (for which $n_{e} = 10^{17}$m$^{-3}$,
$R_{D} = 10^{-9}$m),
and for plasmas induced by lasers in
thermonuclear fusion experiments
(for which $n_{e} = 10^{28}$m$^{-3}$,
$R_{D} = 5 \cdot 10^{-10}$m) \cite{plasma}
eq. (26) gives just $\alpha \cong h$
up at most one order of magnitude.

In the other cases of celestial, gaseous discharge and 
thermonuclear fusion plasmas, which evidently are
in more unstable states with respect to the above
cited plasmas, the order of magnitude of 
$\alpha$ is sensibly greater with respect to 
Planck constant (for each of these systems we
have $\alpha \cong 10^{-29}J\cdot s$).
Finally, two cases (interstellar matter
and solar horn plasmas ) cannot be treated at
all, due to their strong instability
and the consequent large uncertainties 
on the available experimental data.

In conclusion, taking into account
the above exposed difficulties, we can assert
that our method works in a not unsatisfactory
way also for neutral plasmas, again showing
the presence of relevant quantum signatures in
macroscopic systems.

We now show that a much more stringent check
for our scheme can be obtained by studying the
case of charged beams in particle accelerators.
These systems, in fact, are without doubt
mantained, during very long time periods
in a stable and ``coherent" state of motion
by the constant application of focussing
external magnetic fields and by impulsive
energy pumping from radiofrequency cavities.

In the reference frame comoving with the bunch
of the beam, one can consider 
the transverse oscillations of a charged 
particle with respect to 
the ideal circular orbit. Also in this
case the associated effective force is 
in first approximation elastic 
(neglecting in the magnetic field terms
higher then the dipolar contributions):
\begin{equation}
F = -k_{b} x \, ,
\end{equation}

\noindent where $x$ is the transverse displacement 
and $k$ the effective
elastic constant. From eq. (14) we have then
\begin{equation}
\alpha \cong m^{1/2} k_{b}^{1/2} R^{2} \, ,
\end{equation}
\noindent with $m$ the mass of the particles (protons
or electrons) and $R$ the transverse
dimension of the bunch.

If we now insert in eq. (28) the most
recent data from Hera proton accelerator 
and from the electron linear collider,
we have that for the Hera proton accelerator 
$k_{b} \cong 10^{-12} N \cdot$m$^{-1}$,
$R \cong 10^{-7}$m, while for the 
electon linear collider the data are
$k_{b} \cong 10^{-11} N \cdot$m$^{-1}$,
$R \cong 10^{-7}$m.

With these
values, and by inserting respectively proton
and electron masses, eq. (28) yields for
$\alpha$ in both cases
just the order of magnitude of Planck constant. 

Closing this subsection, we may say that although our treatment
based on effective forces is certainly qualitative, 
they yield for the characteristic action $\alpha$ 
the order of magnitude of Planck constant for 
electromagnetic systems on all spatial scale ranges, 
and for a number of charged constituents 
that varies from $N=2$ (hydrogen atom) to macroscopic
values.

\vspace{0.5cm}

{5.3 {\it Strong interactions}}

We now move to consider hadrons 
having as granular constituents collections 
of bound quarks.

The interaction we consider is the ``string law" described by
the typical confining potential $V=k r$, with the string
constant $k$ varying in the range $k \cong 0.1 GeV \cdot fm^{-1}
\div 10 GeV \cdot fm^{-1}$ (values compatible with the 
experimental bounds \cite{perkins}). 

The quark masses vary in the range $m \cong 0.01 GeV \cdot c^{-2}
\div 10 GeV \cdot c^{-2}$ \cite{particledata}, while the
dimension $R$ is the range of nuclear forces, $R \cong 10^{-15} m$.
The constant string force $F=k$, together with eq. (14), gives
for the characteristic action
\begin{equation}
\alpha \cong m^{1/2} R^{3/2} k^{1/2} \, .
\end{equation}

\noindent 
If we insert in eq. (29) the numerical value of $R$, and some 
intermediate values for the mass $m$ and the string constant
$k$, we obtain also in this case $\alpha \cong h$, the Planck
constant.
The natural objection at this point is that, as long as 
we consider systems of bound quarks, since these objects
``live'' in the deep quantum domain, getting Planck constant
is a trivial matter, since such quantity is directly
encoded in the characteristic physical parameters,
in particular in the estimate of the string constant $k$.

We remark however that, besides the fact that the quark
case supplies a necessary check of consistency for
the congruence of our scheme, it is in any
case worth noting the implicit possibility
to construct an effective ``semiclassical"
fluctuative approximation also for quark
systems.

The second comment is on the expression of the characteristic
action $\alpha$ that can be obtained for a
system of confined quarks by allowing the unit of action
to depend on $c$, and therefore by taking into account,
in analogy with electromagnetic systems, the virtual 
exchange of gluons and screening effects.
Following the same line of reasoning adopted in
the previous subsection, we end up then with the
following expression for the unit of action per
constituent:
\begin{equation}
\alpha \cong (mc^{2})^{3/2}c^{-1}k^{-1/2}R^{1/2} \, .
\end{equation}

\noindent Inserting numbers, we obtain again Planck
action constant, up to at most one or two orders of 
magnitude. 
The force law associated to such expression for
$\alpha$ is now modified with respect to the linear
string law, as it acquires a form $\sim 1/R^{2}$.
However, this is not so surprising.
First of all, a similar term must be added to the 
confining one in the phenomenological form of the 
interaction between quarks \cite{perkins} to take
into account the repulsive effects at short distances.
Furthermore, it is possible that also the string--like
contribution is simply an effective description in a stable,
confining region of, for instance, a three--body 
system of quarks, while its real two--body interaction
might always be of the Gauss form $\sim 1/R^{2}$.

We will not dwell deeper into this problem.
However, it is relevant for our general line of
reasoning that, following two different qualitative
procedures, we always obtain $\alpha \cong h$.

\section{Fluctuative hypothesis and temperature:
the equivalent thermal unit of action}

In the picture we have outlined till now, a crucial role is
assigned to the characteristic time $\tau$.
It is then important to estimate its order of magnitude 
for different systems through some independent method.

Since our procedure is of a semiquantitative nature, 
an explicit expression for $\tau$ can provide consistency checks
of the fundamental fluctuative hypothesis, to yield further
support to its universal validity.

It is well known that the typical time scale of quantum fluctuations
can be defined as the ratio between $h$ and a suitable energy describing 
the equilibrium state of a given system about its characteristic
dimensions. This leads naturally to identify this energy with some
equivalent thermal energy $k_{B} T$, with $k_{B}$ the 
Boltzmann constant and $T$ some equivalent temperature
measured in units of the absolute Kelvin scale.

On the other hand, in our scheme such time scale coincides, at
least in order of magnitude, with the fluctuative time $\tau$.
Having tested that it always holds that 
$\alpha \cong h$, we then write
\begin{equation}
\tau \cong \frac{h}{k_{B} T} \, .
\end{equation}

It is interesting to note that, with the above definition, we
obtain from the universal tremor hypothesis eq. (1) the
following suggestive definition of the equivalent temperature:
\begin{equation}
T \cong \frac{h}{k_{B}} \frac{\sqrt{N}}{{\cal{T}}} \, ,
\end{equation}

\noindent which in turn leads to the following
general definition of the equivalent thermal
unit of action:
\begin{equation}
A_{T,N} \equiv (k_{B}T){\cal{T}} \cong h \sqrt{N} \, .
\end{equation}

\noindent We remark the nontrivial nature of this
definition that links the thermal unit of action to
the square root of the number of elementary constituents
through the Planck action constant.

We can apply relations (31)--(33) to some well established
phenomena, as a further test of validity
of our theoretical scheme.
We start by considering charged beams in particle
accelerators.

Among the stable aggregates of charged particles, they play
a paradigmatic role. They are very interesting from our
point of view because, although they are generally described
in classical terms due to their macroscopic dimensions
\cite{fasci}, quantum aspects seem to be however relevant
expecially for electron beams.

Moreover, quantum--like treatments of these systems, attempting
descriptions based on some wave properties of the dynamics at
a ``mesoscopic" level appear very effective and in some cases
also experimentally well verified \cite{quantumlike}.

In a companion paper \cite{noifasci} we introduce
a quantum--like approach to charged beams whose basic
ingredient is a diffusive kinematics ruled by the
effective ``mesoscopic" diffusion coefficient (33), which
simulates semiclassical corrections to classical dynamics
by means of classical stochastic processes.
The quantum--like equations of motion in hydrodynamic
form are then obtained by applying to the 
mean classical stochastic action the well established
techniques of stochastic variational principles 
that have been extensively and successfully developed in
the framework of Nelson stochastic mechanics, an 
independent and self--consistent reformulation
of non relativistic quantum mechanics in the
language of classical stochastic processes
\cite{variational}.

To the end of implementing this program is however 
crucila that we first obtain in our semi--quantitative
framework the correct order of magnitude of the characteristic
physical quantities associated to charged beams. 
The typical quantity we are interested in is the 
thermal beam emittance ${\cal{E}}_{T}$ \cite{fasci}.

The thermal beam emittance is a scale of action 
(or equivalently of temperature, or length) 
which characterizes the spreading in phase space
of the collective motion of a charged particle beam.
It then plays the role of a diffusion
coefficient in the stochastic treatment, and of
a fundamental quantum of action in the quantum--like
description of beams \cite{quantumlike}.

For various electron machines its value in units of the
Compton length $\lambda_{c}=h/mc$ ($m$ the electron mass),
extrapolated from experiments, lies in a range
${\cal{E}}_{T} \cong 10^{6} \lambda_{c} \div 10^{9} \lambda_{c}$
\cite{renato}.

Following our scheme we must assume that the
natural expression for the characteristic thermal action
associated to charged beams be given by eq. (33).
Since in a typical bunch $N \cong 10^{11} \div 10^{12}$
\cite{particledata}, we obtain ${\cal{E}}_{T} \cong 10^{6} 
\lambda_{c} \cong 10^{-6} m$, in good agreement with 
the estimated phenomenological order of magnitude. 
Our scheme shows then in this case a good
degree of consistency.
Again, eq. (33) is remarkable in that it provides an
intriguing relation between the (transverse) emittance
of a charged beam and the number of particles in the beam.
The connection with the square root of $N$ ``summarizes" the
complex fluctuative phenomena taking place in a bunch
of interacting charged particles in accelerators.
As the validity of eq. (33) would imply relevant
consequences for physics of charged beams, it
would be important to supply further theoretical
arguments, or devise experimental tests which could
confirm it.

We now move on to consider another very typical 
macroscopic system such as a gas constitued by 
$N \cong 10^{23} mol^{-1} \div 10^{24} mol^{-1}$
(the Avogadro number) particles.
If we take ${\cal{T}} \cong 10^{-2} s \div 10^{-3} s$,
corresponding to a rms velocity $v_{T} \cong 10^{2} m
\cdot s^{-1} \div 10^{3} m \cdot s^{-1}$ (for gases
at room temperature), we obtain from eq. (32) that
the our equivalent fluctuative temperature takes the
value $T \cong 10^{2} K \div 10^{3} K$, i.e. room 
temperature, as ith should be.

At the opposite end of the ladder, 
we pick up then a typical microscopic
system such as a set of confined 
quarks in a nucleon, for which we can 
naturally assume $N \cong 1$.
The typical energy scale $E$ of light quarks in a
nucleon is of the order of $\Lambda_{QCD} \cong 0.1 GeV$
\cite{particledata}, corresponding to an equivalent 
temperature $T=E/k_{B} \cong 10^{12} K$.
Being the system relativistic, the characteristic
velocity is of the order of the velocity of light $c$.
Since the scale of length is $R \cong 10^{-15} m$, we
have ${\cal{T}} \cong R/c \cong 10^{-23} s$, and with
these numerical values eq. (32) yields exactly the
correct order of magnitude $T \cong 10^{12} K$.

Finally, let us consider a thermodynamic system that
exhibits macroscopic quantum behavior. A typical example
is provided by a Bose condensate.
Bose--Einstein condensation has been recently
experimentally observed in a gas of rubidium and 
sodium \cite{bose}, and then later achived employing
different chemical elements in several
laboratories around the world.
We here use eqs. (31) and (32) 
to estimate the characteristic rms velocity
of an atom in the condensate.

The condensate reaches a linear dimension $R \cong 10^{-4} m$
at a temperature $T \cong 10^{-6} K$, and it contains
$N \cong 10^{7}$ atoms. Letting ${\cal{T}} \cong R/v$, with
$v$ the characteristic velocity, eq. (31) yields
\begin{equation}
v \cong \frac{k_{B}}{h} \frac{T}{\sqrt{N}} R \, .
\end{equation}

\noindent 
Inserting numerical values, $v$ turns out to be of the
order of $10^{-2} m \cdot s^{-1}$
as experimentally observed.

As a last example, we come back to the most macroscopic
system one may imagine, i.e. the whole Universe, and
employ our scheme to estimate the total 
number of nucleons contained in it.
To start with, we interpret, as usually done,
the measured temperature associated to the cosmic 
background radiation, $T=2.7 K$, as the 
characteristic equivalent temperature of the Universe.

We insert in eq. (32) a characteristic global time ${\cal{T}}
\cong R/v$, with $R$ the observed radius of the Universe and
$v$ a characteristic velocity pertaining to 
microscopic constituents of the Universe. 
This velocity can be extimated, in the linear classical regime,
by the equipartion theorem: $mv^2 \cong k_{B}T$, which,
together with eq. (32) provides the following expression
for the total number $N$ of nucleons in the Universe:
\begin{equation}
N \cong m k_{B} T R^{2} h^{-2} \, ,
\end{equation}

\noindent
which, inserting the numerical values that were
previously quoted for the nucleon
mass, the Boltzmann constant, the cosmic backgrounnd
temperature, the observed radius of the Universe and
the Planck action constant, yields $N \cong 10^{78}$.
This result, obtained in our crude qualitative
framework, is in extraordinarily amazing agreement
with the currently accepted central value 
$N \cong 10^{\nu}$, $\nu=78 \pm 8$, 
estimated by cosmological arguments \cite{peebles}.

\section{Conclusions}

By postulating a general principle of mechanical
stability for bound confined dynamical systems,
we have derived the necessity of a classically
chaotic fluctuative component of the classical
motion, first introduced as a hypothesis by
Calogero in the case of gravitational systems.

We have then developed a general semiquantitative
scheme that allows to
compute the order of magnitude of the characteristic
action per constituent in systems ruled by generic,
overall attractive law of forces, including the
crucial instances of electromagnetic and strong 
interactions.

We have shown that in order of magnitude the  
characteristic action per particle is always 
equal to Planck constant for several generic
aggregates with the dimensions of the different aggregates
ranging from microscopic to macroscopic scales, going
from systems of confined quarks in nucleons up to
the whole numbre of gravitationally interacting
nucleons in the Universe (the case first envisaged
by Calogero). 

We have thus shown that the original proposal by
Calogero can be greatly extended and assumes a
universal validity, irrespective of the interactions
and of the systems being considered.
In fact, from the universal validity of the analysis of
Calogero, we infer that quantum mechanics and the
law of force associated to a given confined system
determine the dimensions of coherence 
of any physical system (from quarks in nucleons to the
whole Universe) by imposing a classical principle
of mechanical stability.

Our extension extracts from the original proposal
by Calogero the very interesting and promising 
possibility of obtaining , for any physical system,
fundamental quantum aspects at the classical 
level by introducing proper elementary classical
fluctuations.

This procedure can turn out to be useful in general, and in
particular for the study of phenomena lying at the
boundary between the classical and the
quantum domains \cite{defalco}. 
In these regions it can be
also interesting to explore whether, and to what extent,
Nelson stochastic mechanics, i. e. a
line of thought that associates classical stochastic
processes to the dynamics of quantum states may 
acquire a more fundamental status rather than
being just a mathematical reformulation
of quantum mechanics in terms of classical
stochastic processes \cite{stochmech}.

It is worth noting that, among all the physical systems for which
the scheme has been tested, a particularly relevant
role is played by the macroscopic charged
aggregates which, just as the Universe,
are in fact commonly included in the classical
domain, where Planck constant should not play
any significant role. Viceversa, by remaining
in a purely classical setting, we show that 
quantum mechanics leaves its significant footprint
also on the classical arena in a higly nontrivial way.
In our opinion, it is 
exactly the very crude and semiquantitative
character of our analysis that makes this result
expecially compelling.
Our next effort will be to explicitely construct an
algorithm which allows the present scheme to rest on 
more quantitative grounds.

Coming to terms with the conceptual issues that
our analysis opens wide, the first thought must
concern the physical origin of the coherent tremor,
which remains an open question. 
One might say that our findings support an enlarged 
Calogerian point of view, asserting that a tiny  
chaotic (stochastic) component of the motion 
actually affects the dynamics
of any classical system, with gravity playing no
particular privileged role, and that this universal
chaoticity is at the origin of quantization.
The chaoticity would be universal, in spite of the
differences between, say, the gravitational and the
electromagnetic interactions, because of the scaling
behavior of the couplings and of the spatial ranges of
the different forces, whose ratios are adjusted to yield
$h$ in all instances.

A more cautious and sensible stand is, to us, to 
regard our results as disclosing remarkable methodological 
perspectives, in particular the possibility to formally describe
some relevant aspects of quantum mechanics in 
the mathematical language of classical dynamics complemented
with that of classical stochastic processes.
The universality of the procedure seems to rule
out the (undoubtely suggestive) possibility of a privileged role
of classical gravitation as the ``origin'' of quantum mechanics.

The indication seems rather to be that quantum mechanics 
(whatever the physical origin of it)
and the particular law of force ruling a given physical system 
determine the coherence length of the latter 
under proper conditions of dynamical stability:
remarkably, our analysis shows that this is also 
true for the whole Universe, when treated as an isolated
dynamical system.
The idea that quantum mechanics fundamentally
determines the observed radius of the Universe 
is certainly intriguing {\it per se}. But, more than that,
we stress that the connection between quantum mechanics
and the dimensions of stability of physical systems on any 
arbitrary scale, which is due in many cases to extremely complex 
processes and needs to be explained by extremely sophisticated
formalisms, is obtained in the present context via
a very direct and intuitive line of thought.

For instance,
it is very well known that in a stationary
state the bound electron in the hydrogen atom does not radiate.
This is the great achievement of quantum mechanics in explaining
the stability of matter interacting via electromagnetic forces.
One should anyway be aware of the fact that in classical mechanics
stability of the hydrogen atom is not completely ruled out, once
one realizes that the naive explanations of the classical 
instability (based essentially on the equipartion theorem)
becomes higly questionable once one considers the much
more sophisticated analyses on the complicated nonlinear
interactions of the electron with the infinitely many modes of
the classical electromagnetic field \cite{galgani}.
Now, once we consider the infinetely many modes of the classical
electromagnetic field, or, if one wishes, the infinite
fluctuating number of photons of the quantized electromagnetic
field, the electron--proton system becomes a very complex 
nonlinear dyanmical system. We would then be led to consider 
the possibility of an enlarged Calogerian interpretation
also for electromagnetic and strong interactions, considering
even the most elementary aggregates compatible with such
interactions as truly formed by an enormous amount of
interacting constituents which would exhibit on the proper
space--time scales some classical chaotic behavior of a
possibly coherent nature.

To sum up, as already mentioned the highly remarkable fact is
that the reasoning of Calogero and our analysis open the way
to the definition of an algorithm which starts from the
classical (phenomenological) formulation, and then allows 
room for relevant quantum footprints on the basis of 
a suitable, purely classical fluctuative hypothesis.

We expect that such a theoretical approach,
when stated in a fully quantitative framework,
might turn up to be a powerful computational 
and conceptual tool expecially when applied 
at the borderline 
between the classical and the quantum domain, 
and it is therefore worth of further studying 
and deeper understanding.

\newpage

\section{Aknwoledgements}

We are indebted to Francesco Guerra for the many discussions
that we had with him concerning our work, and for the
precious and far reaching insights he provided us in the
many years of our scientific and personal acquaintance.
His decisive advise and generous help have been
of paramount importance to us at all stages of preparation
of the present paper. 

One of us (F. I.) aknowledges the
Alexander von Humboldt--Stiftung for financial support,
and the Quantum Theory Group of Prof. Dr. Martin Wilkens
at the Fakult\"at f\"ur Physik der Universit\"at Potsdam
for hospitality while on leave of absence from the Dipartimento
di Fisica dell'Universit\`a di Salerno.


\vspace{0.2cm}

\begin{center} {\bf REFERENCES} \end{center}

\begin{thebibliography}{99}

\bibitem{noilettera} S. De Martino, S. De Siena and F. Illuminati,
Mod. Phys. Lett. {\bf B 12}, 291 (1998).

\bibitem{calogero} F. Calogero, Phys. Lett. {\bf A 228}, 335 (1997).

\bibitem{weinberg} S. Weinberg, {\it Gravitation and Cosmology:
Principles and Applications of the General Theory of Relativity}
(Wiley, New York, 1972).

\bibitem{peebles} P. J. Peebles, {\it Principles of Physical Cosmology}
(Princeton University Press, Princeton, N. J., 1993).

\bibitem{landau} L. D. Landau and E. M. Lifshitz, {\it Mechanics}
(Pergamon Press, Oxford, 1969).

\bibitem{plasma} R. J. Goldstone and P. H. Rutherford, {\it
Introduction to Plasma Physics} (Princeton University Press,
Princeton, N. J., 1995).

\bibitem{perkins} D. H. Perkins, {\it Introduction to High Energy Physics}
(Addison Wesley, Menlo Park, $3^{rd}$ edition, 1993).

\bibitem{particledata} A.A. V.V., {\it Review of Particle Physics},
Phys. Rev. {\bf D 54}, 1 (1996).

\bibitem{fasci} S. Chattopadhyay, AIP Conf. Proc. {\bf 127}, 444
(1983); F. Ruggiero, A. Picasso and L. A. Radicati, Ann. Phys. 
(N. Y.) {\bf 197}, 396 (1990).

\bibitem{quantumlike} R. K. Varma, in
{\it Quantum--like Models and Coherence Effects}, R. Fedele and 
S. Shuckla editors (World Scientific, Singapore, 1996), 
and references therein.

\bibitem{noifasci} N. Cufaro Petroni, S. De Martino, S. De Siena
and F. Illuminati, Electronic Preprint physics/9803036 (1998),
to appear in the Proceedings of the International Workshop on
``Quantum Aspects of Beam Dynamics'', held in Stanford, 4--9
January 1998.

\bibitem{variational} F. Guerra and L. M. Morato, Phys. Rev.
{\bf D 27}, 1774 (1983); E. Nelson, {\it Quantum Fluctuations} 
(Princeton University Press, Princeton, N. J., 1985);
F. Guerra and R. Marra, Phys. Rev. {\bf D 29}, 1647 (1984);
L. M. Morato, Phys. Rev. {\bf D 31}, 1982 (1985); F. Guerra,
Ann. Inst. Henri Poincar\'e {\bf 49}, 315 (1988).

\bibitem{renato}  R. Fedele, G. Miele and L. Palumbo, 
Phys. Lett. {\bf A 194}, 113 (1994), and references
therein.

\bibitem{bose} M. H. Anderson {\it et al.}, Science
{\bf 269}, 198 (1995); K. B. Davies {\it et al.}, Phys.
Rev. Lett. {\bf 75}, 3969 (1995).

\bibitem{defalco} D. de Falco, S. De Martino and S. De Siena,
Phys. Rev. Lett. {\bf 49}, 181 (1982); S. De Martino, S. De Siena,
F. Illuminati and G. Vitiello, Mod. Phys. Lett. {\bf B 8}, 977
(1994).

\bibitem{stochmech} F. Guerra, Phys. Rep. {\bf 77}, 263
(1981); S. De Martino, S. De Siena and F. Illuminati,
J. Phys. {\bf A 30}, 4117 (1997); F. Illuminati and
L. Viola, J. Phys. {\bf A 28}, 2953 (1995).

\bibitem{galgani} G. Benettin, L. Galgani and A. Giorgilli,
Nature {\bf 311}, 444 (1984); L. Galgani, A. Giorgilli and
F. Guerra, Phys. Lett. {\bf A 139}, 221 (1989); L. Galgani,
in {\it Probabilistic Methods in Mathematical Physics},
F. Guerra, M. I. Loffredo and C. Marchioro editors
(World Scientific, Singapore, 1992).

\end{thebibliography}
\end{document}